\documentclass[twocolumn,english]{revtex4-2}
\usepackage[T1]{fontenc}
\usepackage[latin9]{inputenc}
\usepackage{amsmath}
\usepackage{graphicx}
\usepackage{babel}
\begin{document}
\title{Jamming on deformable surfaces}
\author{Zhaoyu Xie}
\affiliation{Tufts University, Dept. of Physics \& Astronomy, 574 Boston Ave, Medford,
MA 02155}
\author{Timothy J. Atherton}
\affiliation{Tufts University, Dept. of Physics \& Astronomy, 574 Boston Ave, Medford,
MA 02155}

\begin{abstract}
Jamming is a fundamental transition that governs the mechanical behavior of particulate media, including sand, foam and dense suspensions but also biological tissues: Upon compression, particulate media can change from freely flowing to a disordered solid. Jamming has previously been conceived as a bulk phenomenon involving particle motions in fixed geometries. In a diverse class of soft materials, however, solidification can take place in a deformable geometry, such as on the surface of a fluid droplet or in the formation of a bijel. In these systems the nature and dynamics of jamming remains unknown. Here we propose and study a scenario we call metric jamming that aims to capture the complex interactions between shape and particles. Unlike classical jamming processes that exhibit discrete mechanical transitions, surprisingly we find that metric jammed states possess mechanical properties that are continuously tunable between those of classical jammed and conventional elastic media. New types of vibrational mode that couple particulate and surface degrees of freedom are also observed. Our work lays the groundwork for a unified understanding of jamming in deformable geometries, to exploit jamming for the control and stabilization of shape in self-assembly processes, and provides new tools to interrogate solidification processes in deformable media more generally. 
\end{abstract}

\maketitle

Jamming is a transition to rigidity that occurs as particulate media
are compressed from a freely flowing state to a solid state\citep{van2009jamming,liu2010jamming}.
Jamming can be induced by varying thermodynamic variables such as
temperature and density, as well as mechanical variables such as applied
stress: Colloidal suspensions become colloidal glasses as the density
is increased, flowing foams become static as the shear stress is decreased
below yield stress, liquids become glasses as the temperature is lowered
below the glass transition\citep{liu1998jamming,o2003jamming}. Moreover,
in biological systems, confluent tissues also exhibit a transition
to rigidity controlled by single-cell properties such as shape and
motility\citep{bi2015density,bi2016motility,grosser2021cell,oswald2017jamming}. 

Recently, however, a number of experiments have emerged that involve solidification in deformable geometries and 
do not neatly fit into the jamming scenarios previously envisioned because
these only consider particle-particle interactions and take place
in Euclidean space with a fixed geometry. In these new situations, the rigidification
takes place not only with respect to particle degrees of freedom,
but also with respect to the shape of the interface itself. Further,
the non-Euclidean geometry of the interface means that particles in
different locations may experience location dependent states of stress
depending on the local shape of the interface.

In this work, we therefore propose a new jamming class, which we refer to as \emph{metric
jamming}, that refers to structures jammed both with respect to particle
degrees of freedom and surface degrees of freedom. The purpose of
this paper is to construct a model metric jamming process, explicitly
test the resulting structures for rigidity and hence distinguish similarities
and differences from other jammed media. 

The jammed state has unique properties compared to normal crystalline
solids and new physics emerges near the jamming transition\citep{o2003jamming,van2009jamming}.
In contrast to crystalline solids, jammed materials generally lack
translational order and are fragile, offering little or no resistance
to shear deformation, and exhibit other unusual elastic properties
if the particles themselves are deformable. The fragility arises from
the packing's \emph{isostaticity}, i.e. they possess only the minimal
number of contacts per particle required for mechanical stability.
Understanding jamming of disordered systems will help the
fabrication of new functional amorphous materials\citep{stokes2008rheology,mattsson2009soft,vlassopoulos2014tunable}. 

Jamming theory has been extended to consider consider nonspherical\citep{vanderwerf2018hypostatic,yuan2019jammed} and deformable
\citep{treado2021bridging,wang2021structural} particles as well as
the role of friction\citep{bi2011jamming}. In such extensions, both
\emph{hypo}static and \emph{hyper}static configurations---those with
an apparent deficit or surplus of contacts relative to the isostatic
value---can emerge requiring sophisticated approaches to constraint
counting\citep{henkes2016rigid,papanikolaou2013isostaticity} and
new universality classes\citep{brito2018universality}. 

Two distinct theoretical approaches to jamming have been proposed:
One considers jammed states of soft particles that interact via a
short range interparticle potential $V(r)$\citep{o2003jamming}.
The energy of the system is expanded as a quadratic form $U\sim\delta x\cdot H\cdot\delta x$
about a candidate jammed state of interest, where $H$ is the Hessian
matrix of the energy, also known as dynamical matrix\citep{wyart2005geometric,wyart2005effects,charbonneau2016universal},
and $\delta x$ is the displacement vector. Eigenanalysis of $H$
is used to test the overall stability of the structure, and hence
if it is truly jammed, identify particles (known as \emph{rattlers})
that are superfluous to the rigidity. The spectrum of $H$ also provides
the density of vibrational states and therefore determines the elastic
response. Jammed structures are found to possess an excess of low-frequency
modes that are a signature of the fragility of the state\citep{ghosh2010density,chen2010low,o2003jamming,wyart2005effects,wyart2005geometric,silbert2005vibrations,silbert2009normal,degiuli2014effects,manning2015random,charbonneau2016universal,arceri2020vibrational}. 

An alternative approach considers configurations of rigid, mutually
impenetrable particles under confinement\citep{torquato2010jammed}.
Particles in a candidate structure are subjected to set of random
forces $\xi$ to find a protoypical unjamming motion $\delta x$ identified
by extremization of the virtual work $\xi\cdot\delta x$, a linear
programming problem, subject to (linearized) interpenetrability constraints\citep{donev2004jamming,donev2004linear}.
Analysis of the prototypical unjamming motion leads to a hierarchical
classification of jammed structures\citep{torquato2001multiplicity}:
A packing is \emph{locally jammed}, the least stringent category,
if no particles are able to move while the others remain fixed; it
is \emph{collectively jammed} if no subset of particles is movable
with the remainder held in place; it is \emph{strictly jammed} if
no collective subset of the particles can be moved at the same time
as a volume conserving deformation of the container.

Examples of solidification in deformable geometries include the phenomenon of arrested coalescence
in Pickering emulsions\citep{pawar2011arrested}, mixtures of immiscible
fluids stabilized by the addition of colloidal particles adsorbed
at the fluid-fluid interface of the constituent droplets. Here, emulsion
droplets successively coalesce, leading to gradual densification of
the particles and, ultimately, rigid arrested structures that inhibit
further coalescence. Deformation of Pickering emulsion droplets by
an electric field can rigidify in wrinkled patterns\citep{mikkelsen2019mechanical}.
Another example involves colloidal particles immersed in a host fluid
that undergoes a phase transition to a liquid crystalline phase; the
moving phase boundary drives the particles to remain in the vanishing
isotropic phase which densify and jam forming rigid shells\citep{Rodarte2013}.
Other situations involve more exotic particles, such as bacteria on
the interface of an oil-in-water emulsion droplet that they gradually
consume\citep{dorobantu2004stabilization}, or non-compact interfaces,
such as the the production of jammed emulsion gels (bijels) from bicontinuous
precursor mixtures\citep{stratford2005colloidal,herzig2007bicontinuous}.

\section*{Model system}

To do so, we consider the following simplified scenario: suppose $N$
soft spherical frictionless particles with different radii $r_{i}$
are positioned with their centroids at coordinates $\mathbf{X}_{i}$
on a closed compact surface $\partial C$ that bounds a region $C$
of fixed volume representing, for example, an emulsion droplet. To
facilitate comparisons with studies of jamming in flat space\citep{o2003jamming,gao2006frequency,silbert2009normal,xu2005random,vaagberg2011glassiness,goodrich2012finite,chen2018stress,boromand2019role},
we will use $50-50$ mixtures of bidispersed particles with radius
ratio $1:1.4$ to suppress crystallization. For simplicity, we assume
that the particles are rigidly confined to the surface and that their
presence does not significantly deform the interface locally by forming
mensici. The particles interact with one another through a potential
of finite range $V(d_{ij})$ where $d_{ij}$ is the separation of
particles $i$ and $j$.
\begin{figure}
\begin{centering}
\includegraphics[width=3.5in]{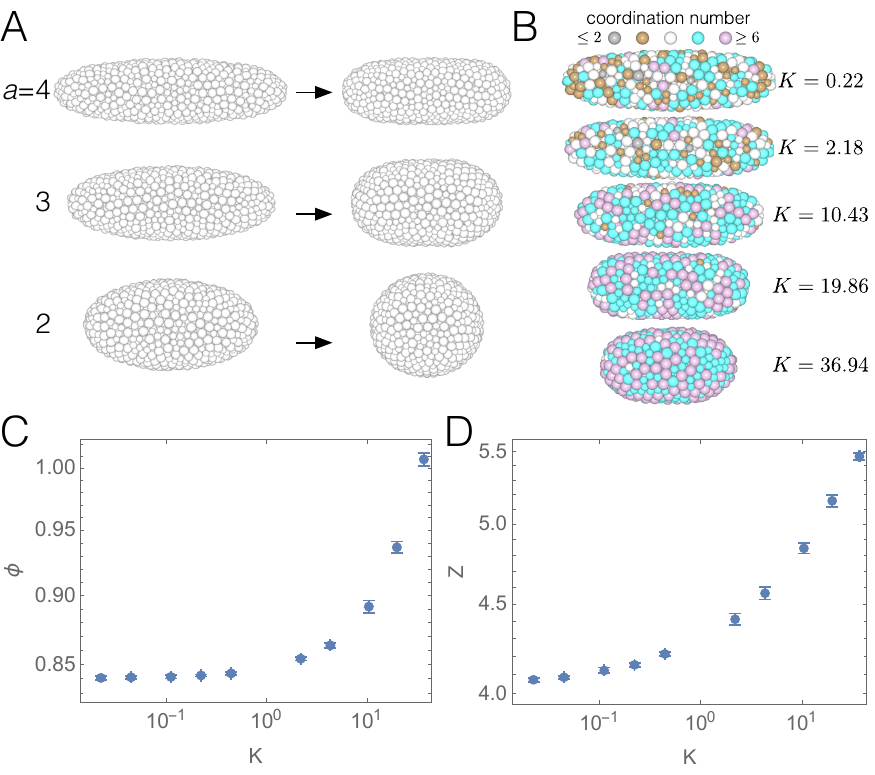}
\par\end{centering}
\caption{\textbf{\label{fig:Metric-jammed-configurations.}Metric jammed configurations.
A} Initially jammed configurations on ellipsoids of varying aspect
ratio (left) are relaxed into metric jammed structures (right). \textbf{B}
Final metric jammed configurations as a function of $K$. \textbf{C}
Packing fraction as a function of $K$. \textbf{D} Mean contact number
per particle as a function of $K$.}

\end{figure}

The total energy of the system includes both surface tension and particle-particle
interactions,

\begin{equation}
E=\sigma\int dA+k\sum_{i\neq j}V(d_{ij}),\label{eq:totalEnergy}
\end{equation}
where $\sigma$ is the surface tension and $k$ is the rigidity of
the particles. A dimensionless ratio $K=\sigma A/k$ characterizes
the relative importance of these terms where $A$ is the surface area.
The choice of surface energy here is specific to the Pickering emulsion
scenario described above: The integral in (\ref{eq:totalEnergy})
is over the exposed fluid-fluid interface of the emulsion droplet
which might be locally deformed due to capillary effects. For simplicity,
we will here assume that the particles each occupy a fixed interfacial
area and instead integrate over the area of the whole surface, supplementing
(\ref{eq:totalEnergy}) with a constant $-\sum_{i}\pi r_{i}^{2}$
that does not enter into our later calculations where $r_{i}$ is
the radius of a single particle. Other surface energies might be applicable
in different scenarios, e.g. the Helfrich energy for jamming on a
vesicle\citep{helfrich1973elastic}, for example, or an elastic energy
for a membrane.

The surface is parameterized by a map $\mathbf{X}(\mathbf{x})=\mathbf{x}R(\mathbf{x})$
from points $\mathbf{x}$ that lie on the unit sphere to 3D Euclidean
space where the radial function $R(\mathbf{x})$ is decomposed into
tesseral harmonics, 
\begin{equation}
R(c_{lm},\mathbf{x})=\sum_{lm}c_{lm}Z_{lm}(\mathbf{x}).\label{eq:tesseralHarmonics}
\end{equation}
The configuration of the system may be fully specified by the set
of parameters $\mathbf{\mathbf{\xi}}=\{\mathbf{x}_{i},c_{lm}\}$ including
$N$ particle positions $\mathbf{x}_{i}$ on the unit sphere and $M$
surface coefficients $c_{lm}$ for a total of $2N+M$ degrees of freedom.
Fixing the volume enclosed by the surface imposes a nonlinear constraint
on the surface coefficients $\{c_{lm}\}$ and removes one degree of
freedom. The physical position $\mathbf{X}_{i}$ of the $i$th particle
may be calculated from the map $\mathbf{X}$ and depends on $\mathbf{x}_{i}$
and $\{c_{lm}\}$.

For particle-particle interactions we use a compact repulsive pairwise
potential\citep{o2003jamming,wyart2005effects,silbert2005vibrations,xu2005random,gao2006frequency,xu2010anharmonic,vaagberg2011glassiness,goodrich2012finite,charbonneau2016universal,chen2018stress,boromand2019role,vanderwerf2020pressure},

\begin{equation}
V(d_{ij})=\frac{1}{2}\left(1-\frac{d_{ij}}{s_{ij}}\right)^{2}\Theta\left(\frac{d_{ij}}{s_{ij}}-1\right),\label{eq:hertianSphere-1}
\end{equation}
where the separation $d_{ij}=\left|\mathbf{X}_{i}-\mathbf{X}_{j}\right|$,
$s_{ij}=r_{i}+r_{j}$ and $\Theta$ is the Heaviside step function
enforcing interaction only for overlapping particles.

\section*{Simulations}

We construct metric jammed configurations by the following procedure:
we first create rigid packings on surfaces of fixed shape, specifically
ellipsoids of varying aspect ratio. These will be used to help distinguish
the physical consequences of the curved geometry from those associated
with the surface modes. Using the ellipsoidal packings as a starting
point, we then minimize the total energy (\ref{eq:totalEnergy}) of
the configuration with respect to both particle and surface parameters
$\xi$, producing a final jammed structure. Further details are provided
in the Methods section below. Fig. \ref{fig:Metric-jammed-configurations.}A
displays initial and final states for several different aspect ratios. In the majority of the paper, we use $N=800$ particles, a figure large enough to mitigate finite size effects but sufficiently small for computational expediency. 

For each configuration, the stability is assessed by calculating and
diagonalizing the bordered hessian of the energy functional (\ref{eq:totalEnergy})
with respect to the parameter set $\xi$ incorporating both particle
and surface degrees of freedom and including the volume constraint.
Particle coordinates are parametrized in spherical polar form $\mathbf{x}_{i}=R(\sin\theta_{i}\cos\phi_{i},\sin\theta_{i}\sin\phi_{i},\cos\theta_{i})$.
Rattlers, particles that do not contribute to the rigidity of the
structure, are identified from zero modes of the hessian associated
with eigenvectors that are localized to a single particle; these particles
are then removed from the structure. There also exist zero modes associated
with residual degrees of freedom; for ellipsoids there is one such
mode associated with cylindrical symmetry about the long axis; for
spherical packings there are three. In practice, the zero modes arising
from numerical calculations mix combinations of rattler and trivial
motions which must be separated by Gram-Schmidt orthogonalization
of the associated eigenvectors. 

\section*{Structural analysis}

Metric jammed structures for different values of $K$ for $N=800$ particles are shown in
Fig. (\ref{fig:Metric-jammed-configurations.})B. For large $K$,
with large surface tension, the final shape tends towards spherical,
and hence achieves a globally minimal surface; for smaller $K$, or
larger inter-particle interaction energies, the shape instead tends
toward spherocylindrical. From these, we can compute two structural
measures, the packing fraction $\phi=(N\pi/A)\sum_{i}r_{i}^{2}$,
where $A$ is the area of the surface at the jamming point, and the
mean number of contacts per particle $Z$, that are signatures of
jamming in flat space. Fig. (\ref{fig:Metric-jammed-configurations.})C
displays $\phi$ as a function of $K$ for $N=400$ particles: As $K\to0$, $\phi$ approaches
the value of $0.84$ characteristic of random close packing in 2D
space\citep{o2003jamming}. As $K$ increases, the particles are increasingly
compressed and the exposed surface gradually eliminated hence $\phi\to1$
for large $K$. The mean contact number $Z$ is plotted with respect
to $K$ in Fig. (\ref{fig:Metric-jammed-configurations.})D also for $N=400$ particles and tends
towards $4$ as $K\to0$ which for $K\to10$ the number of contacts
is significantly greater, reaching values as high as $5.5$.

We now compare these results with the situation in flat space. A linear
constraint counting argument predicts the minimal number of contacts:
the number of degrees of freedom for $N$ particles is $dN$ where
$d$ is the dimensionality and $\xi$ is the number of residual degrees
of freesom in the space ($\xi=d$ in $d$-dimensional space and $\xi=1$
for cylindrically symmetric surfaces as discussed above). These must
be balanced by $NZ_{iso}/2$ constraints, i.e. $dN=NZ_{iso}/2+\xi$.
The isostatic contact number in $2$D is therefore $Z_{iso}=4-4/N$.
Previous literature demonstrates that an additional degree of freedom
is required to maintain positive pressure and bulk modulus in flat
space, such that $dN+1=NZ_{c}+\xi$\citep{goodrich2012finite,goodrich2014jamming}.
Thus the contact number at jammed state $Z_{c}=4-2/N$ on 2D flat
surfaces and $Z_{c}=4$ on cylindrically symmetric surfaces. Hence,
the metric jammed structures produced are generally \emph{hyperstatic}
and reproduce the isostaticity observed in flat space only as $K\ll1$,
i.e. where the rigidity of the particles is significantly greater
than the surface tension. In that limit, the packing fraction approaches
the random close packing value.

\begin{figure}
\begin{centering}
\includegraphics[width=3.5in]{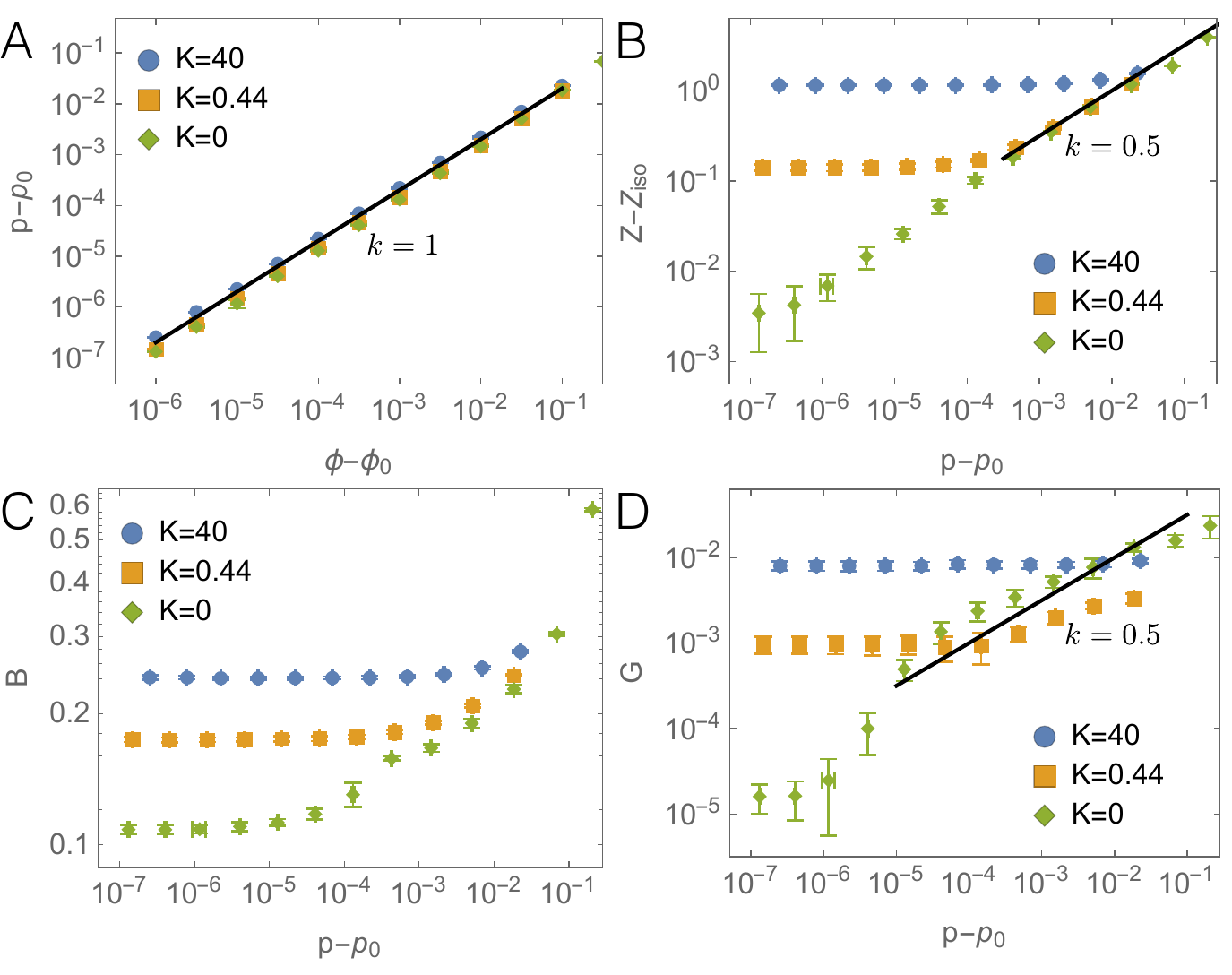}
\par\end{centering}
\caption{\textbf{\label{fig:Mechanical-properties-of}Mechanical properties
of metric jammed configurations. }(a) Relation between pressure and
packing fraction. (b) change in contact number as a function of pressure.
Elastic moduli as a function of pressure for different $K$: (c) bulk
modulus and (d) shear modulus.}
\end{figure}

\section*{Mechanical Properties}

Jammed structures in 2D flat space also exhibit a unique elastic response:
they possess vanishing shear modulus, and exhibit characteristic scalings
of the pressure $p\sim(\phi-\phi_{0})^{1}$, excess contact number
$\Delta Z\equiv Z-Z_{iso}\sim(\phi-\phi_{0})^{1/2}$, bulk and shear
elastic moduli $B\sim(\phi-\phi_{0})^{0}$, $G\sim(\phi-\phi_{0})^{1/2}$
as the system is compressed beyond the jamming point $\phi_{0}$\citep{o2003jamming,liu2010jamming,goodrich2012finite,goodrich2014jamming}.
To test this, we deform metric jammed configurations in two ways:
the structure is compressed slightly to measure the bulk modulus $B$
and a twist deformation is imposed about the symmetry axis to measure
the shear modulus $G$. The elastic moduli are computed from derivatives
of the stress tensor during the deformation as described in Methods
and results for packings with different $K$ are displayed in Fig.
\ref{fig:Mechanical-properties-of}.

Fig. \ref{fig:Mechanical-properties-of}(a), shows that the excess
pressure $\Delta p=p-p_{0}$ is proportional to the excess packing
fraction $\phi-\phi_{0}$ and is therefore a good measure of the deformed
system's proximity to the jamming point. The excess contact number,
bulk modulus and shear modulus are displayed as a function of $\Delta p$
in Fig. \ref{fig:Mechanical-properties-of}(b)-(d). The excess coordination
number displays an initial plateau but then increases at a critical
pressure that increases with $K$. Far from the jamming point, the
excess contact number scales as $\Delta Z\sim(\Delta p)^{0.5}$. The
bulk modulus possesses a similar plateau for all $K$ at low $\Delta p$.
The shear modulus $G$ is several orders of magnitude smaller than
the bulk modulus, and increases approximately proportionally to $K$.
For small $K$, $G$ scales approximately like $(\Delta p)^{0.5}$;
this behavior is reproduced for larger $K$ only at higher pressure
where a plateau exists for small $\Delta p$. Hence, these results
well reproduce the behavior of jammed packings in flat space at low
$K$. As $K\gg1$, the structures begin to resemble normal solids
where $G\sim(\phi-\phi_{c})^{0}$\citep{o2003jamming,liu2010jamming,goodrich2012finite,goodrich2014jamming}
even though $G$ remains small. 

\begin{figure*}
\begin{centering}
\includegraphics[width=5.5in]{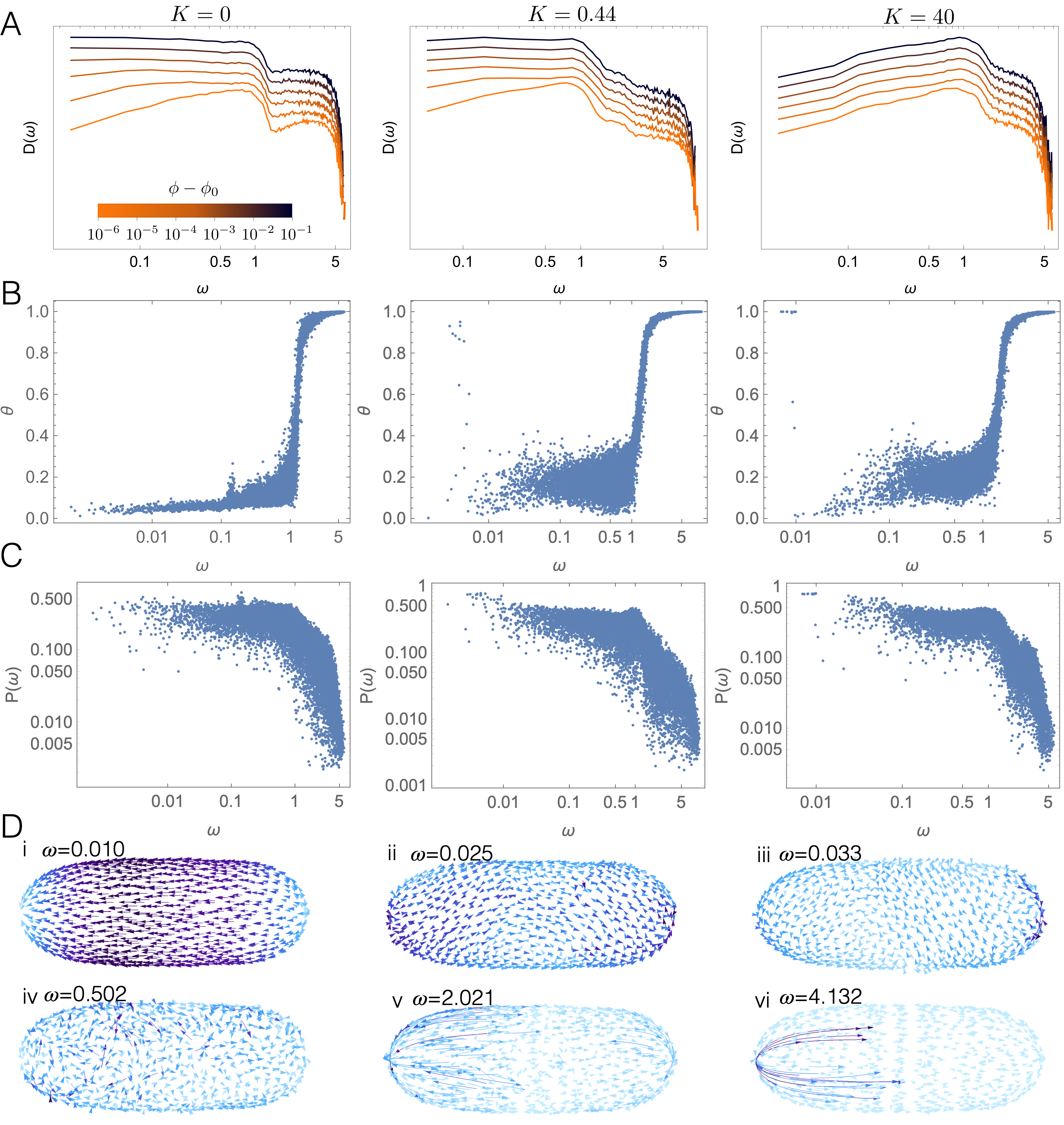}
\par\end{centering}
\caption{\label{fig:Eigenanalysis.}\textbf{Vibrational properties of metric
jammed configurations.} \textbf{A} Vibrational density of states $D(\omega)$,
\textbf{B} vibrational component along polar axis for $\phi-\phi_{0}=10^{-6}$
and \textbf{C} participation ratio $P(\omega)$ for metric jammed
configuraion with $K=0$, $K=0.44$ and $K=40$ at $\phi-\phi_{0}=10^{-6}$.
\textbf{D} Selected vibrational modes for $K=40$: \textbf{i} translation
along polar direction, \textbf{ii} phonon mode, \textbf{iii} quasi-localized
mode, \textbf{iv} extended mode along polar direction, \textbf{v}
extended mode along azimuthal direction, \textbf{vi} localized mode.}
\end{figure*}

\section*{Vibrational modes}

A third signature of the jammed state is an excess of low frequency
vibrational modes\citep{o2003jamming,wyart2005effects,wyart2005geometric,silbert2005vibrations,silbert2009normal,chen2010low,degiuli2014effects,charbonneau2016universal,mizuno2017continuum,wang2019low,arceri2020vibrational},
in contrast to the Debye theory of elastic solids\citep{debye1912theorie,zeller1971thermal}.
Eigenanalysis of the hessian matrix provides the vibrational frequencies,
which are the square roots of the eigenvalues measured in units of
$\sqrt{k}/r$ where $r$ is the radius of the larger particle, and
associated vibrational modes from the eigenvectors. We display the
density of states $D(\omega)$ for different values of $K$ in Fig.
\ref{fig:Eigenanalysis.}A at several values of $\Delta\phi=\phi-\phi_{0}$.
Two key features are apparent: first, $D(\omega)$ possesses a low
frequency plateau characteristic of jamming for low $K$ and at low
$\Delta\phi$. As the system is compressed, or as $K$ increases,
the plateau vanishes leading to Debye-like behavior.

A second feature is that $D(\omega)$ falls abruptly above a critical
value of $\omega$. The transition is very sharp as $K\to0$ and smoothed
out for larger $K$. In Fig. \ref{fig:Eigenanalysis.}B, we display
the component of the eigenvectors in the polar direction as a function
of $\omega$, close to the metric jammed configuration. At low frequency,
particles tend to vibrate along the azimuthal $\phi$ direction while
high-frequency localized vibrations are along the polar $\theta$
direction. The crossover of $D(\omega)$ coincides with the transition
of particle motions. For packings on spherical surfaces the high frequencies
have motions both along $\theta$ and $\phi$ directions due to the
symmetric nature. Hence spatially inhomogeneous curvature can cause
the localization of vibrational modes.

The participation ratio $P(\omega)$ is calculated to characterize
the vibrational modes\citep{chen2010low,xu2010anharmonic,charbonneau2016universal,mizuno2017continuum,arceri2020vibrational}.
It is a measure of the fraction of particles that are participating
in the motion governed by the mode of frequency $\omega$. Given the
eigenvectors $\left\{ \overrightarrow{u}_{i}(\omega)\right\} $ at
frequency $\omega$ for every particle, 
\begin{equation}
P(\omega)=\frac{1}{N}\frac{(\sum_{i}|u_{i}(\omega)|^{2})^{2}}{\sum_{i}|u_{i}(\omega)|^{4}},\label{eq:participationRatio}
\end{equation}
where $N$ is the number of particles after removing ratters. On the
curved surfaces considered here, we use the arclength to represent
$|u_{i}(\omega)|^{2}$.

The corresponding participation ratios $P(\omega)$ are also displayed
in Fig. \ref{fig:Eigenanalysis.}C together with illustrations of
typical vibrational modes in Fig. \ref{fig:Eigenanalysis.}D for configurations
close to the metric jamming point with $K=40$, revealing the consequences
of anisotropic curvature. In the low-frequency region far from the
jamming point, plane-wave like phonon modes (Fig. \ref{fig:Eigenanalysis.}D
ii) and quasi-localized modes (Fig. \ref{fig:Eigenanalysis.}D iii)
where localized excitations are visible on a small plane-wave background.
In the mid-frequency region, we have extended modes where random excitations
spread throughout the entire system along the azimuthal $\phi$ direction
(Fig. \ref{fig:Eigenanalysis.}D iv), and along the polar $\theta$
direction (Fig. \ref{fig:Eigenanalysis.}D v) for larger frequencies.
The high-frequency region contains localized modes where only a few
particles vibrate along the polar $\theta$ direction(Fig. \ref{fig:Eigenanalysis.}D
vi). These vibrational modes also exists for jammed structures on
flat surfaces\citep{zeravcic2008localization,van2009jamming,silbert2009normal,liu2010jamming,mizuno2017continuum,arceri2020vibrational}.
The spatially inhomogeneous curvature, however, causes the extended
modes along the polar $\theta$ or azimuthal $\phi$ direction. Additionally
for metric jammed configurations at very low frequency where the component
of eigenvectors along $\theta$ direction and participation ratio
are almost $1$, there appears a vibrational mode where all particles
move along the $\theta$ direction, showed in Fig. \ref{fig:Eigenanalysis.}i.

\begin{figure}
\begin{centering}
\includegraphics[width=3.5in]{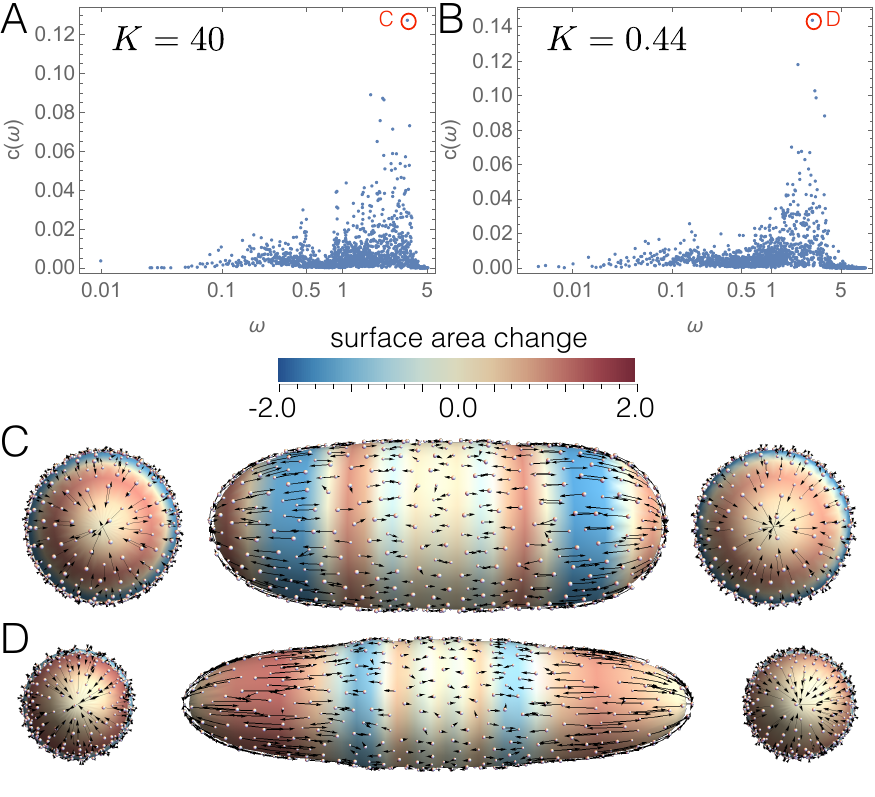}
\par\end{centering}
\caption{\label{fig:SurfaceMode}\textbf{Combined shape-particle vibrational
modes.} \textbf{A} Projection $c(\omega)$ of the eigenvector associated
with the shape deformation sector for rigid particles $K=40$ and
\textbf{B} $K=0.44$. \textbf{C} and \textbf{D} Representative eigenvectors
with significant shape and particle deformation; the local rate of
change in area is plotted together with arrows indicating particle
motions. }
\end{figure}

Our inclusion of shape as well as particulate degrees of freedom allows
for the possibility of new modes with mixed character. To examine
this, we display in Fig. \ref{fig:SurfaceMode}A and B the projection
of each normalized eigenvector that lies in the shape sector, $c(\omega)$,
as a function of the angular frequency. We see that modes associate
with significant shape deformation tend to lie in the high frequency
regime, relatively independently of $K$. Two such modes are depicted
in Fig. \ref{fig:SurfaceMode}C and D.

\section*{Conclusion}
These results collectively show that particulate media on deformable
surfaces can form structures that share structural and mechanical
properties with conventional jammed media in Euclidean space, but
are rigid with respect to surface as well as particle degrees of freedom.
Experimentally produced structures are likely to initially be arrested,
i.e. only locally jammed, but by successive unjamming and relaxation
events proceed toward a new \emph{metric jammed} state that is rigid
with respect to surface as well as particle degrees of freedom. The
presence of a surface energy tends to compress the particles somewhat
in the final state, and hence metric jammed structures resemble those
in flat space only in the limit of high particle rigidity. By adjusting
the relative influence of surface and particle energies characterized
by a single dimensionless parameter $K=\sigma A/k$, metric jammed
structures can continuously be tuned from isostatic with vanishing
shear modulus, i.e. similar to jammed states in flat space, to states
that resemble conventional elastic solids. A simple structural metric,
the coordination number, serves as an indicator of where in this range
a particular candidate structure lies. The vibrational spectrum may
similarly be tuned from similar to a jammed solid, possessing an excess
of low frequency modes, to Debye-like. In either case, the curved
space leaves a significant imprint on the spectrum, leading to localized
and oriented modes due to the anisotropic curvature.

Our work provides a theoretical framework that unifies our understanding
of solidification processes that take place on deformable media and
extends the applicability of the jamming concept. Beyond the Pickering
emulsion example developed here in detail that can be directly realized
experimentally, the results may provide fresh insight into other materials
such as bijels\citep{stratford2005colloidal,herzig2007bicontinuous}
that rely on such processes. A particularly exciting potential class
of applications lies in biological matter, given the remarkable success
in applying ideas from jamming to tissues and particularly tumor progression\citep{bi2015density,bi2016motility,oswald2017jamming,grosser2021cell}.
Future work should aim to integrate other developments from the jamming
community, such as nonspherical particles and deformability\citep{grosser2021cell,treado2021bridging,brito2018universality,vanderwerf2018hypostatic},
to elucidate possible connections between the shape of the confining
surface and particle shape. While the configurations observed here
were hyperstatic, the nontrivial coupling of surface and particle
degrees of freedom, and the new kinds of deformation mode that can
emerge, forshadows the recently identified need for more sophisticated
approaches to characterizing rigidity\citep{damavandi2021energetic1}.
Experimentally, the interplay of order and shape that emerges in metric
jamming suggests the possibility of assembly processes that sculpt
particulate media into a desired configuration by exploiting deformable
interfaces; our framework provides a unified understanding to facilitate
their design. 

\acknowledgements{We thank P. Spicer and A. Donev for helpful discussions. This material is based upon work supported by the National Science Foundation under Grant No. DMR-1654283.}

\section*{Materials \& Methods}

\subsection*{Jammed configurations on fixed surfaces}

We adapt the protocol for the generation of jammed configurations
in 2D and 3D space\citep{o2003jamming} to packings on curved surface
using the interparticle potential shown in Eq. (\ref{eq:hertianSphere-1}).
The simulation starts by randomly placing $N$ particles of potentially
different radii $r_{i}$ with their centroids fixed on a curved surface
that is scaled such that the area $A\gg N\pi r^{2}$ and hence the
packing fraction $\phi\ll1$. The simulation proceeds by iteratively
reducing the size of the surface at fixed shape to slowly increase
the packing fraction. Conjugate gradient descent is applied at each
iteration to bring the particle interaction energy to the minimum;
the sequence is terminated if the total potential energy per particle
$V/N<10^{-16}$ or $V/N$ for successive iterations deviates by less
than $10^{-15}$. As the size of the surface decreases and the particles
move closer to each other, inevitable overlaps appear and increase
the energy minimum. The algorithm is halted if the minimized total
potential energy per particle falls into $10^{-16}<V/N<2\times10^{-16}$.
This procedure brings the system extremely close to temperature $T=0$,
with a very small pressure $p<10^{-10}$ as calculated below. These
thresholds give a clear separation between jammed and unjammed state\citep{vaagberg2011glassiness}
and a similar approach has been utilized in other research about packings
\citep{xu2005random,gao2006frequency,mailman2009jamming,schreck2010comparison,xu2010anharmonic}.
Packings are explicitly tested for rigidity by eigenanalysis as described
in the main text. 

\subsection*{Production of metric jammed configurations}

We begin with a jammed packing on a fixed surface produces as described
in the preceding section. Given such a configuration, described by
parameters $\xi$, generalized forces are evaluated by taking the
gradient of the energy functional (\ref{eq:totalEnergy}), including
the surface energy, with respect to the surface coefficients $\mathbf{f}_{c_{lm}}=-\nabla_{c_{lm}}E$.
The surface is deformed along the descent direction with volume conservation
by a stepsize $\delta$ and then the conjugate gradient method is
employed to bring the particle system back to an energy minimum, as
discussed above. The deformation is accepted if the total energy of
the system afterwards is reduced, otherwise the deformation is rejected
and the stepsize reduced $\delta\to\delta/2$. Further surface deformation
steps are taken and the algorithm is stopped if $\delta<10^{-16}$
or energy $E$ for successive iterations deviates by less than $10^{-12}$.

\subsection*{Calculation of mechanical properties}

We calculate the stress tensor at each particle by constructing a
local frame with tangent vectors $\mathbf{t}_{\theta}$ and $\mathbf{t}_{\phi}$
aligned in the polar and azimuthal directions respectively. The stress
tensor in 2D\citep{allen2017computer,o2003jamming,schreck2010comparison,goodrich2016scaling,chen2018stress,boromand2019role}
is then written as, 
\begin{equation}
\Sigma_{\alpha\beta}=\frac{1}{A}\sum_{i>j}(r_{ij\alpha}f_{ij\beta}+r_{ji\alpha}f_{ji\beta})/2,\label{eq:stressTensorCurv}
\end{equation}
 where $A$ is the surface energy and $r_{ij\alpha}$, $f_{ij\beta}$
represent the projections of center-to-center distance and force along
surface tangent vectors $\mathbf{t}_{\theta}$ and $\mathbf{t}_{\phi}$.
The average is computed over pairs of particles. From this, we can
compute the pressure $p=(\Sigma_{\theta\theta}+\Sigma_{\psi\psi})/2$,
bulk modulus $B=\psi dp/d\psi$ after slightly compressing the system
at packing fraction $\phi$ and shear modulus $G=-d\Sigma_{\theta\psi}/d\gamma$
after applying a small shear strain $\gamma$\citep{o2003jamming,chen2018stress}.
These variables are measured in units of $k/r^{2}$ where $r$ is
the radius of the larger particle. The shear is applied by twisting
the configuration around the ellipsoid symmetry axis, 
\[
\psi_{i}=\begin{cases}
\begin{array}{c}
\psi_{i}+2\theta_{i}\gamma\\
\psi_{i}+2(\theta_{i}-\pi)\gamma
\end{array} & \begin{array}{c}
0\leq\theta_{i}<\pi/2\\
\pi/2<\theta_{i}\leq\pi
\end{array}\end{cases}.
\]
after which we apply the conjugate gradient descent method to minimize
the energy while fixing the position of several particles near the
poles; these fixed particles and the area they cover are excluded
from the stress tensor calculation.

\bibliographystyle{rsc}
\bibliography{ref,newref}

\end{document}